# PREVIRIALIZATION: PERTURBATIVE AND N-BODY RESULTS

E. L. Łokas[1], R. Juszkiewicz[1,2], F. R. Bouchet[3] and E. Hivon[3]

[1] Copernicus Astronomical Center, Bartycka 18, 00-716 Warsaw, Poland
[2] Institute for Theoretical Physics, University of California,
Santa Barbara, California 93106-4030, USA
[3] Institut d'Astrophysique de Paris, 98 bis Bd Arago, 75014 Paris, France

E–mail: lokas@camk.edu.pl (ELL), roman@camk.edu.pl (RJ), bouchet@iap.fr (FRB),
hivon@iap.fr (EH)

## ABSTRACT

We present a series of N-body experiments which confirm the reality of the previrialization effect. We also use weakly nonlinear perturbative approach to study the phenomenon. These two approaches agree when the rms density contrast, $\sigma$, is small; more surprisingly, they remain in agreement when $\sigma \approx 1$. When the slope of the initial power spectrum is $n > -1$, nonlinear tidal interactions slow down the growth of density fluctuations and the magnitude of the suppression increases when $n$ (i.e. the relative amount of small scale power) is increased. For $n < -1$ we see an opposite effect: the fluctuations grow more rapidly than in linear theory. The transition occurs at $n = -1$ when the weakly nonlinear correction to $\sigma$ is close to zero and the growth rate is close to linear. Our results resolve recent controversy between two N-body studies of previrialization. Peebles (1990) assumed $n = 0$ and found strong evidence in support of previrialization, while Evrard & Crone (1992), who assumed $n = -1$, reached opposite conclusions. As we show here, the initial conditions with $n = -1$ are rather special because the nonlinear effects nearly cancel out for that particular spectrum. In addition to our calculations for scale-free initial spectra, we show results for a more realistic spectrum of Peacock & Dodds (1994). Its slope near the scale usually adopted for normalization is close to $-1$, so $\sigma$ is close to linear. Our results retroactively justify linear normalization at $8h^{-1}$ Mpc, while also demonstrating the danger and limitations of this practice.

*Subject headings:* cosmology: theory – galaxies: clustering – galaxies: formation – large–scale structure of universe

# 1 Introduction

It is generally believed that the structures observed in the universe today grew by gravitational instability out of much smaller inhomogeneities present at the epoch of the decoupling of matter and radiation. The details of this general picture, however, are still under discussion. One of the sources of disagreement is the size of the so called previrialization effect. The term "previrialization" was introduced by Davis & Peebles (1977), although the general idea was expressed earlier (Peebles & Groth 1976).

According to the previrialization conjecture, small scale fluctuations present within collapsing mass concentrations tend to slow down their growth. Initial asphericities and tidal interactions between neighboring density fluctuations induce significant nonradial motions, which oppose the collapse. Virialized clumps form later than predicted by frequently used approximations, such as linear perturbation theory or the spherical collapse solution (another way to state the problem is to say that to obtain a given final density contrast, in case of a realistic mass distribution, one needs a significantly larger initial contrast than the one for an isolated spherical fluctuation). Some N-body experiments (Villumsen & Davis 1986; Peebles 1990) appear to reproduce these effects, while others do not. An example of the latter is the paper by Evrard & Crone (1992).

Using his least action method of evolving particle orbits back in time, Peebles (1990) compared histories of isolated clumps with that of an ensemble of mass concentrations. The required initial density contrast in the former case was consistent with the spherical model. The initial contrast required for the more realistic mass distribution was significantly larger. Although the spherical model predicts that the growth of the density contrast is faster than the linear approximation, the growth observed in the numerical action method is slower.

Evrard & Crone (1992) posed the question, whether small-scale structure affects the clustering on larger scales. Their N-body simulations supported the conclusion that the abundance of rich clusters finally formed was insensitive to the amount of small-scale power present in the initial conditions. One of our purposes here is to address this controversy.

Interactions between fluctuations at different scales can be also studied using weakly nonlinear perturbation theory. A formalism, appropriate for this purpose, was proposed over a decade ago (Peebles 1980; Juszkiewicz 1981; Vishniac 1983; Fry 1984; Juszkiewicz, Sonoda & Barrow 1984). This formalism uses Newtonian hydrodynamics and an important assumption that the linear term in the perturbative expansion for the density contrast is a Gaussian random field. Unfortunately, N-body simulations available in the early eighties were too crude to study the range of validity of perturbative methods. This made further progress difficult and the field became dormant for several years. With time, the dynamical range of the simulations improved. Moreover, in the last decade the depth of galaxy surveys has increased by an order of magnitude, allowing accurate measurements of spatial correlations of galaxies at separations large enough to make weakly nonlinear theory useful in their analysis.

As a result of these developments, the study of the evolution of the mass distribution in the expanding universe has become a rich topic for analytic perturbative methods. The renewed enthusiasm for analytic calculations produced many new results (for a review, see e.g. Bouchet & Juszkiewicz 1993; Sahni & Coles 1995; Juszkiewicz & Bouchet 1996 and references therein). Those directly relevant for our purposes here are contained in the papers of Suto & Sasaki (1991) and Makino, Sasaki & Suto (1992). Studying the nonlinear evolution of the power spectrum, these authors succeeded in reducing the dimensionality of the mode coupling integrals, which appear in the formalism of Vishniac (1983) and Juszkiewicz et al. (1984). As a result, they were able to obtain closed-form expressions for spectral corrections for scale-free initial conditions. Using the same formalism, Jain & Bertschinger (1994) investigated the transfer of power in the CDM spectrum. They compared their results to N-body simulations, finding a good agreement in a wide dynamical range.

In this paper, we address the previrialization controversy using weakly nonlinear perturbation theory. As a measure of the previrialization effect we calculate the weakly nonlinear corrections to the second moment (the variance) of the cosmic density field. As it was already pointed out by Juszkiewicz et al. (1984), and confirmed by Suto & Sasaki (1991), Makino et al. (1992), and Jain & Bertschinger (1994), the strength and the wavenumber dependence of nonlinear corrections to the power spectrum are determined by the shape of the initial spectrum. Hence, we may expect that the strength of the previrialization effect may also vary with initial conditions. As we will show here, this is indeed the case and the differences in conclusions, reached by Peebles (1990) and Evrard & Crone (1992) may actually reflect the differences in the slopes of the power spectra in their simulations.

This paper is organized as follows. In section 2, we summarize the formalism of weakly nonlinear perturbation theory and calculate the second order contributions to the variance. We use Gaussian initial conditions and perform the calculations for scale-free spectra and a more realistic spectrum of Peacock & Dodds (1994), based on data from several galaxy surveys. The results and discussion are given in Section 3, where we check our perturbative calculations against a set of N-body simulations. We also compare perturbation theory with calculations, based on a phenomenological model of nonlinear clustering, similar to that of Hamilton et al. (1991).

## 2   Perturbation theory

We assume a model universe with vanishing cosmological constant and arbitrary density parameter $\Omega$. The content of the universe is supposed to behave as a pressureless fluid which undergoes gravitational evolution described by the usual Newtonian equations. The cosmic density field is characterized by the density contrast $\delta = \delta(\mathbf{x}, t) = \delta\rho/\rho_b$, where $\mathbf{x}$ is the Eulerian comoving coordinate, $t$ is the cosmological time, $\rho_b$ denotes the background density, and $\delta\rho$ is the difference between local mass density and $\rho_b$.

The perturbative expansion of the density contrast around the background solution $\delta = 0$ is
$$\delta = \delta_1 + \delta_2 + \delta_3 + \cdots \tag{1}$$
where $\delta_n = \mathcal{O}(\delta_1^n)$, $\delta_1$ being the linear theory solution. The $n$-th order solutions are obtained from equations describing the Newtonian evolution using the solutions of the $(n-1)$-th order of density and velocity fields as source terms (see Fry 1984; Goroff et al. 1986).

In the Einstein-de Sitter universe the scale factor scales with time as $a \propto t^{2/3}$ and the background density $\rho_b = 3\dot{a}^2/8\pi G a^2 = 1/6\pi G t^2$. The time dependence of the $n$-th order follows
$$\delta_n(\mathbf{x}, t) = [D(t)]^n \delta_n(\mathbf{x}) \tag{2}$$
where $D(t) \propto a(t)$ and we consider only the mode growing in time. For an arbitrary cosmological model, however, the time dependences of different orders should be considered independently. Fortunately for a wide range of $\Omega$ the solutions for the density contrast are very weakly dependent on $\Omega$ and the Einstein-de Sitter case provides a good approximation (Bouchet et al. 1992, 1995; Catelan et al. 1995).

All of the following calculations are much simpler if they are performed in Fourier space. For the first order of the density contrast field we have
$$\delta_1(\mathbf{k}, t) = D(t) \int \mathrm{d}^3 x \ \delta_1(\mathbf{x}) \mathrm{e}^{-i\mathbf{k}\cdot\mathbf{x}} \tag{3}$$
and the inverse Fourier transform is
$$\delta_1(\mathbf{x}, t) = D(t)(2\pi)^{-3} \int \mathrm{d}^3 k \ \delta_1(\mathbf{k}) \mathrm{e}^{i\mathbf{k}\cdot\mathbf{x}}. \tag{4}$$

For the following calculation only second and third order solutions for the density contrast are needed, and we give them here in the Fourier representation (e.g. Goroff et al. 1986).
$$\delta_2(\mathbf{k}, t) = \frac{D^2}{(2\pi)^3} \int \mathrm{d}^3 p \int \mathrm{d}^3 q \ \delta_D(\mathbf{p}+\mathbf{q}-\mathbf{k})\delta_1(\mathbf{p})\delta_1(\mathbf{q}) P_2^{(s)}(\mathbf{p}, \mathbf{q}) \tag{5}$$

$$\delta_3(\mathbf{k}, t) = \frac{D^3}{(2\pi)^6} \int d^3p \int d^3q \int d^3m \; \delta_D(\mathbf{p}+\mathbf{q}+\mathbf{m}-\mathbf{k}) \delta_1(\mathbf{p}) \delta_1(\mathbf{q}) \delta_1(\mathbf{m}) P_3^{(s)}(\mathbf{p}, \mathbf{q}, \mathbf{m}). \quad (6)$$

The symbol $\delta_D$ represents the Dirac delta. The symmetrized kernels are of the form

$$P_2^{(s)}(\mathbf{p}, \mathbf{q}) = \frac{1}{14} J(\mathbf{p}+\mathbf{q}, \mathbf{p}, \mathbf{q}) \quad (7)$$

$$\begin{aligned} P_3^{(s)}(\mathbf{p}, \mathbf{q}, \mathbf{m}) = \; & A[\; H(\mathbf{p}+\mathbf{q}+\mathbf{m}, \mathbf{p}) J(\mathbf{q}+\mathbf{m}, \mathbf{q}, \mathbf{m}) + \\ & + \; H(\mathbf{p}+\mathbf{q}+\mathbf{m}, \mathbf{q}+\mathbf{m}) \, L(\mathbf{q}+\mathbf{m}, \mathbf{q}, \mathbf{m})] + \\ & + B \;\; F(\mathbf{p}+\mathbf{q}+\mathbf{m}, \mathbf{p}, \mathbf{q}+\mathbf{m}) L(\mathbf{q}+\mathbf{m}, \mathbf{q}, \mathbf{m}) + \\ & + \begin{pmatrix} \mathbf{p} \to \mathbf{q} \\ \mathbf{q} \to \mathbf{m} \\ \mathbf{m} \to \mathbf{p} \end{pmatrix} + \begin{pmatrix} \mathbf{p} \to \mathbf{m} \\ \mathbf{q} \to \mathbf{p} \\ \mathbf{m} \to \mathbf{q} \end{pmatrix} \end{aligned} \quad (8)$$

where $A = 1/108$ and $B = 1/189$. In the expression above the notation follows that of Makino et al. (1992) i.e.

$$H(\mathbf{p}, \mathbf{q}) = \frac{\mathbf{p} \cdot \mathbf{q}}{q^2} \quad (9)$$

$$F(\mathbf{p}+\mathbf{q}, \mathbf{p}, \mathbf{q}) = \frac{1}{2} \frac{|\mathbf{p}+\mathbf{q}|^2 \mathbf{p} \cdot \mathbf{q}}{p^2 q^2} \quad (10)$$

$$J(\mathbf{p}+\mathbf{q}, \mathbf{p}, \mathbf{q}) = 4 \frac{(\mathbf{p} \cdot \mathbf{q})^2}{p^2 q^2} + 7 \frac{p^2 + q^2}{p^2 q^2} \mathbf{p} \cdot \mathbf{q} + 10 \quad (11)$$

$$L(\mathbf{p}+\mathbf{q}, \mathbf{p}, \mathbf{q}) = 8 \frac{(\mathbf{p} \cdot \mathbf{q})^2}{p^2 q^2} + 7 \frac{p^2 + q^2}{p^2 q^2} \mathbf{p} \cdot \mathbf{q} + 6. \quad (12)$$

The smoothing of the fields is introduced by the convolution of the density and the filtering function

$$\delta_R(\mathbf{x}, t) = \int d^3y \, \delta(\mathbf{y}, t) \, w(|\mathbf{x}-\mathbf{y}|, R) \, . \quad (13)$$

We use lower-case $w$ for spatial filters and capital $W$ for their Fourier representation, given by

$$W(k) \equiv \int d^3x \, w(x) e^{-i\mathbf{k}\cdot\mathbf{x}} \, . \quad (14)$$

Our window functions are normalized so, that

$$\int d^3x \, w(x) \; = \; W(0) \; = \; 1 \, . \quad (15)$$

We perform our calculations for two types of windows – a Gaussian and a top-hat with Fourier representations respectively

$$W_G(kR) = e^{-k^2 R^2 / 2} \quad (16)$$

and
$$W_{TH}(kR) = 3\sqrt{\frac{\pi}{2}}(kR)^{-\frac{3}{2}}J_{\frac{3}{2}}(kR) \tag{17}$$

where $J_{\frac{3}{2}}$ is a Bessel function

$$J_{\frac{3}{2}}(x) = \sqrt{\frac{2}{\pi x}}\left(\frac{\sin(x)}{x} - \cos(x)\right) . \tag{18}$$

We assume a Gaussian distribution for $\delta_1$ in equation (1) and define

$$\sigma^2 = \langle \delta_1^2 \rangle = D^2(t) \int \frac{d^3k}{(2\pi)^3} P(k) W^2(kR) \tag{19}$$

as the linear variance of the density field. We assume that for $\sigma < 1$, the first few terms in the perturbative expansion provide a good approximation of the exact solution (this assumption is supported by results of N-body simulations, discussed below). Since $\delta_1$ is assumed to be a Gaussian random field, all its statistical properties as well as those of the higher order terms in the perturbative series (1) are determined by the power spectrum $P(k)$, defined as

$$\langle \delta_1(\mathbf{k})\delta_1(\mathbf{q}) \rangle = (2\pi)^3 \delta_D(\mathbf{k} + \mathbf{q}) P(k) . \tag{20}$$

We begin by considering spectra with a power-law form

$$P(k) = Ck^n, \quad -3 \leq n \leq 1 , \tag{21}$$

where $C$ is a normalization constant. For such fields, smoothed with a Gaussian filter, the linear order contribution to the variance (equation 19) is

$$\sigma_G^2 = CD^2(t)\frac{\Gamma(\frac{n+3}{2})}{(2\pi)^2 R^{n+3}} . \tag{22}$$

For a top-hat smoothing we get

$$\sigma_{TH}^2 = CD^2(t)\frac{9\Gamma(\frac{n+3}{2})\Gamma(\frac{1-n}{2})}{8\pi^{3/2}R^{n+3}\Gamma(1-\frac{n}{2})\Gamma(\frac{5-n}{2})} . \tag{23}$$

The perturbative series (1) can be used to expand the second moment of the density field in powers of $\sigma$,

$$\langle \delta^2 \rangle = \langle \delta_1^2 \rangle + \langle \delta_2^2 \rangle + 2\langle \delta_1 \delta_3 \rangle + \ldots \equiv \sigma^2 + (I_{22} + I_{13})\sigma^4 + \mathcal{O}(\sigma^6). \tag{24}$$

The first term is the linear variance given by equation (19). The two following terms are the lowest order nonlinear corrections. These terms are of order $\sigma^4$ because $\delta_1$ is a Gaussian random field (all odd-order terms in the expansion, including those of order $\sigma^3$, must vanish). We have introduced the quantities $I_{22}$ and $I_{13}$ because it is convenient to calculate the first nonlinear corrections in a normalized form, i.e. divided by $\sigma^4$. Indeed, provided the integrals are convergent, for the scale-free power-law spectra, these ratios should be dimensionless numbers, independent of scale, as in the case of the higher moments (e.g. the skewness and the kurtosis) of the fields. By using the second and third order solutions (5) and (6) we obtain

$$I_{22} = \frac{D^4(t)}{98(2\pi)^6 \sigma^4} \int d^3 p \int d^3 q \ P(p)P(q)W^2(|\mathbf{p}+\mathbf{q}|R) \ J^2(\mathbf{p}+\mathbf{q},\mathbf{p},\mathbf{q}) \qquad (25)$$

$$\begin{aligned} I_{13} &= \frac{6D^4(t)}{(2\pi)^6 \sigma^4} \int d^3 p \int d^3 q \ P(p)P(q)W^2(qR) \\ &\times \ \{ A[\ H(\mathbf{q},-\mathbf{p}) \ J(\mathbf{p}+\mathbf{q},\mathbf{p},\mathbf{q}) \\ &\quad +H(\mathbf{q},\mathbf{p}+\mathbf{q}) \ L(\mathbf{p}+\mathbf{q},\mathbf{p},\mathbf{q})] \\ &\quad -B\ F(\mathbf{q},-\mathbf{p},\mathbf{p}+\mathbf{q}) \ L(\mathbf{p}+\mathbf{q},\mathbf{p},\mathbf{q})\} + \begin{pmatrix} \mathbf{p} \to \mathbf{q} \\ \mathbf{q} \to \mathbf{p} \end{pmatrix} \end{aligned} \qquad (26)$$

where the last term in brackets means that similar expression with $\mathbf{p}$ and $\mathbf{q}$ interchanged should be added if the symmetry in $\mathbf{p}$ and $\mathbf{q}$ is to be maintained. After integration the expressions become

$$I_{ij} = \frac{D^4(t)}{2\pi^2 \sigma^4} \int_0^\infty dk \ k^2 \ W^2(kR) P_{ij}(k) \ , \qquad (27)$$

where

$$P_{22}(k) = \frac{k^3}{98(2\pi)^2} \int_0^\infty dx P(kx) \int_{-1}^{+1} d\mu P\left(k\sqrt{1+x^2-2\mu x}\right) \frac{(3x+7\mu-10x\mu^2)^2}{(1+x^2-2x\mu)^2} ; \qquad (28)$$

$$P_{13}(k) = \frac{k^3 P(k)}{252(2\pi)^2} \int_0^\infty dx \ P(kx)$$

$$\times \left[ \frac{12}{x^2} - 158 + 100 x^2 - 42 x^4 + \frac{3}{x^3}(x^2-1)^3(7x^2+2) \ln\frac{1+x}{|1-x|} \right] . \qquad (29)$$

The window function in equation (27) describes spatial smoothing, applied to the density field; each term in the perturbative expansion (1) was convolved with a Fourier transform of $W(kR)$.

The input from $I_{22}$ is always positive and from $I_{13}$ negative, therefore we may interpret their values at some scale respectively as the additional power coming in from other wavelengths and as the power that is lost and taken over by other wavelengths.

Both $I_{22}$ and $I_{13}$ diverge individually in the limit of $k \to 0$ and $kx \equiv q \to 0$ if $n \leq -1$. Fortunately, in the entire interval $-3 \leq n \leq 1$, all diverging terms cancel when $I_{22}$ and $I_{13}$ are added together, so that their sum,

$$I_2 \equiv I_{22} + I_{13} \qquad (30)$$

remains finite. These "miraculous cancellations" are probably associated with the Galilean invariance of the Newtonian equations of motion (Scoccimarro & Frieman 1995). In the opposite limit of $k \to \infty$, $q \to \infty$, divergencies occur for $I_{22}$ if $n > \frac{1}{2}$ and for $I_{13}$ if $n > -1$. Therefore if we want to calculate the sum of both terms for $n > -1$ we need to introduce a cutoff in the initial power spectrum at large wave-numbers

$$P(k) = \begin{cases} Ck^n & \text{for } 0 < k < k_c \\ 0 & \text{for } k > k_c \end{cases} \qquad (31)$$

An equivalent way to introduce the cutoff is to smooth the initial power spectrum with a Gaussian

$$P(k) = Ck^n \, e^{-k^2 r^2} . \qquad (32)$$

We introduce shortwave cutoffs in the power spectrum to avoid divergencies in perturbation theory for scale-free spectra. Diverging terms in the perturbative expansion do not necessarily imply that the fully nonlinear solution for $\sigma(R,t)$ is divergent; strong nonlinearities may "naturally" introduce a nonperturbative cutoff. Moreover, spectra without a cutoff represent a rather far-fetched idealization anyway. In any realistic situation we do expect deviations from pure scale invariance induced by damping processes during the decoupling era (e.g. frozen amplitudes in the radiation era, neutrino free streaming or photon diffusion in the case of Silk damping). In numerical simulations the cutoff is introduced either by the grid or discreetness effects in gridless N-body simulations. In the latter case, $1/k_c$ or $r$ in equation (32) would correspond to the mean interparticle separation (the particles Nyquist scale). In any case, all this suggests that scale-free power spectra are just convenient analytical approximations and the need for a cutoff in some perturbation theory applications may just tell us when this approximation has to break down. Of course, in a real case, the regularization may come by a smooth bending on the spectra rather than a sharp cutoff.

A cutoff of this kind affects the initial conditions for the field $\delta_1$, i.e. the properties of the density field itself. In contrast, spatial smoothing applied at late times, as in equation (27), describes a particular way of measuring the variance, $\langle \delta^2 \rangle$. These two kinds of window functions should not be confused with each other. In linear perturbation theory, for a given choice of a window and a power spectrum, there is no difference between a field that has been spatially smoothed at some initial time and then evolved, and a field which was evolved first and smoothed later. This is no longer true when weakly nonlinear effects are taken into account: smoothing and evolving do not commute. To keep the distinction, we reserve a capital $R$ for the comoving smoothing radius of a filter applied at late times, while lower-case letters, $r$ and $k_c$, for cutoffs present in the initial conditions.

## 3 Results and discussion

### 3.1 Scale-free power spectra

Consider a power spectrum of the form given by equation (31). In the case of no smoothing ($R = 0$, $W = 1$), the linear variance and the nonlinear correction term both diverge when the cutoff is removed by setting $k_c = \infty$. However, when the correction term and the linear variance are calculated for a finite $k_c$, their ratio, $I_2$, remains finite in the limit $k_c \to \infty$. By numerically integrating $I_{22}$ and $I_{13}$ with $W(kR) \equiv 1$, we obtain

$$I_2 = 1.83 \ . \tag{33}$$

This value is very weakly *dependent* on $n$ (contrary to the lowest order values of normalized cumulants). It decreases slightly with growing spectral index $n$, but differences in the range $-3 \le n \le 1$ are smaller than 1 %. Therefore we may approximate the weakly nonlinear $\langle \delta^2 \rangle$ as

$$\sigma_{wnl}^2 = \sigma^2 + 1.83 \, \sigma^4. \tag{34}$$

The above relationship is in excellent agreement with a recent result, obtained independently by Scoccimarro and Frieman (1995), who succeeded in evaluating the integrals $I_{22}$ and $I_{13}$ analytically, and obtained

$$I_2 = I_{22} + I_{13} = \frac{4007}{2205} \approx 1.82 \ . \tag{35}$$

This expression ignores a small $n$-dependent term, which can affect the sum of $I_{22}$ and $I_{13}$ by no more than 1% in the entire range $-2 \le n \le 2$ (again, in agreement with our results of numerical integration).

As we have no other way of checking whether the perturbative series does converge to the true solution, it is important to compare perturbative results with N-body simulations. For the results in equations (34) and (35) this is not possible because they assume no smoothing ($R = 0$), while any experimental measurements in N-body simulations, as well as in galaxy surveys, necessarily involve spatial smoothing over some finite scale $R \ne 0$.

In order to account for spatial smoothing, we have calculated numerically $I_2 \equiv I_{22} + I_{13}$ for a range of spectral slopes, using a Gaussian filter (eq. [16]). Figure 1 shows the results of the calculation of the weakly nonlinear corrections $\langle \delta_2^2 \rangle + 2\langle \delta_1 \delta_3 \rangle$, divided by $\sigma_G^4$. We used power-law spectra with a shortwave cutoff at $k = k_c$ (eq. [31]) (therefore the value of $\sigma_G^2$ is not the one given by eq. [22] but smaller, dependent on the cutoff). The corrections are shown for different values of the dimensionless variable $k_c R$. We see that in the limit $k_c \to \infty$, the integrals are convergent only when $n = -2$. However, for any finite value of $k_c R$, their sum, $I_2$ remains finite, and it is easy to estimate its value analytically for $k_c R \gg 1$. Indeed, we can use the closed form expressions for $P_2(k) = P_{22} + P_{13}$, obtained by Makino et al. (1992), who studied quadratic corrections for power-law spectra with shortwave cutoffs. (Although their formulas are not free from errors, the mistakes we identified do not affect the results presented below.) One can then identify the terms, which dominate in the limit $k_c \to \infty$, substitute the result in equation (27), and then integrate term by term. For $n = -2$, the appropriate expansion for $P_2$ is

$$P_2(k) = \frac{C^2}{(2\pi)^2} \left[ \frac{55\pi^2}{196 k} + \mathcal{O}(1/k_c) \right] . \tag{36}$$

This gives

$$I_2 = \frac{55\pi}{196} \approx 0.882 \tag{37}$$

for a Gaussian window, and

$$I_2 = \frac{1375}{1568} \approx 0.877 \tag{38}$$

for a top-hat. The same approach can be used for the remaining cases, considered in Figure 1. Upon expanding $P_2(k)$ in powers of $k_c$ and neglecting the terms of order $1/k_c$ and higher, one has

$$P_2(k) = \frac{C^2}{(2\pi)^2} \left[ \frac{122}{315} k \ln(k/k_c) + \frac{6136}{11025} k + \mathcal{O}(1/k_c) \right] \qquad (n = -1) ; \tag{39}$$

$$P_2(k) = \frac{C^2}{(2\pi)^2} \left[ -\frac{122}{315} k^2 k_c + \frac{10\pi^2}{147} k^3 + \mathcal{O}(1/k_c) \right] \qquad (n = 0) ; \tag{40}$$

$$P_2(k) = \frac{C^2}{(2\pi)^2} \left[ -\frac{61}{315} k^3 k_c^2 + \frac{9}{49} k^4 k_c - \left( \frac{4973}{44100} + \frac{8}{105} \ln(k/k_c) \right) k^5 + \mathcal{O}(1/k_c) \right] \qquad (n = 1) . \tag{41}$$

Substituting above expansions into equation (27) with a Gaussian window, and integrating termwise, we finally get

$$I_2(k) = 0.638 - 0.387 \ln(k_c R) \qquad (n = -1) ; \tag{42}$$

$$I_2(k) = 1.710 - 0.656 \, k_c R \qquad (n = 0) ; \tag{43}$$

$$I_2(k) = -0.321 + 0.457 \ln(k_c R) + 0.610 \, k_c R - 0.387 \, (k_c R)^2 \qquad (n = 1) \, . \qquad (44)$$

For $k_c R \geq 2$, all of the above asymptotic expressions for $I_2$, as well as the estimate of $I_2$ for $n = -2$ in eq. (37), are in excellent agreement with the results of numerical integration, shown in Figure 1.

Let us now discuss the limits of applicability of the second order perturbation theory as a function of $k_c R$. We must be particularly careful in the limit of large $k_c R$, since $\langle \delta^2 \rangle = \sigma^2 \left[ 1 + \sigma^2 I_2 \right]$, while

$$|I_2| \approx (0.6 \, k_c R)^{n+1} \qquad \text{for} \qquad n > -1 \quad \text{and} \quad k_c R \gg 1 \, . \qquad (45)$$

Hence, the term quadratic in variance could "overtake" the linear term, causing a breakdown of perturbation theory. From these considerations we expect that the approximate limit of validity of the second order results on a scale $R$ is given by the condition

$$\sigma(R) \, < \, |I_2|^{-1/2} \, . \qquad (46)$$

The upper limit on $k_c R$ is therefore

$$k_c R \, < \, (R/R_{nl})^{(n+3)/(n+1)} \qquad (n > -1) \, , \qquad (47)$$

where $R_{nl}$ is the filtering scale, for which the rms density fluctuation equals unity: $\sigma(R_{nl}) = 1$. As we have seen, the results for $n = -2$ are not sensitive to the upper cutoff $k_c$, so the limit of their validity is given by the usual condition $\sigma < 1$ – a standard requirement for perturbation theory. Unless $k_c R$ is "exponentially large", the same limit should apply for the $n = -1$ case, since the divergence in this case is only logarithmic. However, for $n > -1$, the condition (46) becomes more and more restrictive with growing $n$. In particular, we expect that the range for acceptable values of $\sigma$ for $n = 1$ is narrower than the range for $n = 0$.

To check the self consistency of our results, we will now consider the limit $R \to \infty$ for a fixed upper cutoff $k_c$ and a fixed nonlinear scale $R_{nl}$. We should recover the linear perturbation theory, which is indeed the case for $n > -1$ :

$$\langle \delta^2 \rangle \, \approx \, \sigma^2 \left[ 1 - \frac{(0.6 k_c)^{n+1} R_{nl}^{n+3}}{R^2} \right] \, \to \, \sigma^2 \, . \qquad (48)$$

For $-3 < n \leq -1$ the convergence to linear theory is even faster.

In the opposite limit, when $R = 0$, we expect to recover the result for $I_2$ without smoothing, since $W(0) = 1$. Indeed, for $k_c R = 0$, all curves in Fig. 1 do converge on $I_2 = 1.83$, in agreement with equation (34).

The ultimate test for perturbative calculations is the comparison with N-body experiments. We will make such a comparison in the next section. Meanwhile, let us summarize the physical implications of equations (37) and (42) – (44). According to these results, weakly nonlinear effects accelerate the growth rate of density fluctuations when $n = -2$: the rms density contrast grows faster than predicted by linear perturbation theory. However, when there is more power at small scales in the initial spectrum, the nonlinear effects work in the opposite direction and the growth is slowed down. For $n = -1$, the quadratic correction to linear variance is close to zero – the rate is in almost perfect agreement with linear theory. For $n > -1$, the growth rate is slower than the linear prediction. The quadratic correction has a negative sign and its absolute value appears to increase with increasing $n$ as $(k_c R)^{n+1}$, at least for the cases $n = 0$ and $n = 1$, considered here. This behaviour is in qualitative agreement with the previrialization conjecture (Peebles & Groth 1976; Davis & Peebles 1977).

### 3.1.1 Comparison with N-body simulations

To compare predictions of perturbation theory to the N-body results in the case of spectral indices $n = -1, n = 0$ and $n = +1$ we use the simulations made by David Weinberg that were used to check the perturbative calculation of skewness and kurtosis (Juszkiewicz et al. 1995, Lokas et al. 1995). All the simulations used a $200^3$ force mesh and $100^3$ particles (except for n=+1 ones which had $200^3$ particles). The moments of the evolved density field were computed for Gaussian smoothing lengths $L/50$, $L/25$ and $L/12.5$, $L = 100$ cells being the size of the simulation box.

Figure 2 compares the N-body results to the perturbative calculations. Open symbols show the ratios of the N-body nonlinear variance to its linear counterpart calculated from equation (22) using the normalization of the initial power spectra (i.e. $\sigma = 1$ for the final expansion factor $a = 1$ and smoothing scale $L/50$). Circles, triangles and squares correspond to the shortest, medium and largest smoothing scale respectively. The error bars of the results (not plotted) coming from statistical averaging over eight independent simulations are large enough (especially in the $n = -1$ case where the points are most scattered) so that the results do not contradict self-similarity.

A direct, quantitative comparison between the perturbative and N-body results for $n > -2$ is restrained by the difficulty of determining the real small scale cutoff in the initial power spectra of the simulations. The degree of self-similarity displayed by the N-body results would indicate that the cutoff scale is very small. We performed perturbative calculations of the weakly nonlinear corrections to the variance for the spectrum (32) with a Gaussian cutoff. The effect of a Gaussian cutoff in $k$-space is similar to that of a sharp cutoff at $k = k_c \approx 1/r$ (eq. [32]). We chose three different cutoff wavelengths $r$: $r_{Ny}$, $r_{Ny}/2$ and $r_{Ny}/4$, where $r_{Ny}$ is the scale corresponding to the Nyquist frequency. The Nyquist scale is $L/50$ for $n = -1$ and $n = 0$ simulations and $L/100$ in the case of $n = +1$. The choice of a cutoff scale $r$ obviously breaks the self-similarity of the results. We showed in Figure 2 only the results of perturbative calculation corresponding to the medium final smoothing radius $L/25$ (filled triangles). We find that the perturbative results match the N-body best when we assume the cutoff scales $r = r_{Ny}/4$ for $n = -1$, $r = r_{Ny}/2$ for $n = 0$ and $r = r_{Ny}$ for $n = +1$. These values correspond to $k_c R = 8$ for $n = -1$ and $k_c R = 4$ for $n = 0$ and $n = 1$. Using equations (42), (45) and (46), we get $\sigma < 1$, $\sigma < 0.6$ and $\sigma < 0.4$ as approximate limits of validity of perturbation theory for $n = -1, 0$ and $+1$, respectively. These limits, based on the comparison of the linear and quadratic terms, are in reasonable agreement with the true limits of validity, implied by Figure 2.

In the $n = -2$ case, we performed a new simulation with $256^3$ particles (about 17 million) using a PM code (Bouchet, Adam & Pellat 1985; Moutarde et al. 1991) with $256^3$ cells. The initial conditions were imprinted using Zeldovich approximation on a "glass"-like particle distribution (White 1994), with a power equal to 1/25 of the shot noise level at the Nyquist frequency of the particle grid. This "glass" distribution was obtained by N-body simulation in an expanding universe with a *repulsive* gravitation, starting from a random initial distribution. After a high enough expansion (about $10^6$ in our case), particles settle down in a quasi-equilibrium state. This state shows a very uniform particle load without any anisotropy or discernible order down to very small scales. This homogeneity allows a very accurate study of structure formation in high density region, as well as in voids. The variance of the evolved density field was computed on the discreet distribution of a $64^3$ particle sub-sample convolved with a spherical top-hat of radius $R$. Poisson noise associated with discreteness effects has been removed from measured variances.

Figure 3 compares the perturbative predictions and the N-body results for $n = -2$. Open symbols show the ratios of the nonlinear variances to the linear ones, at different expansion factors and at 8 scales, starting from $R = 10^{-2.2}$ times the box size, and spaced by 0.2 in log. The linear variances were calculated according to equation (23) and the perturbative weakly nonlinear approximations to the nonlinear ones (filled symbols in Figure 3) using equation (38). The N-body results closely follow theoretical predictions and obey a self-similar evolution, apart from the largest scale measurements (with $R = 10^{-0.8}$ times the box size). These deviations are likely to be due to the effect of the missing waves at scales larger than the numerical box. The absence of these long waves can also explain the slight offset of the numerical results vs. the theory, because according to Figure 1, for $n = -2$, the correction to linear theory for variance is actually quite sensitive to long waves contribution. We have calculated the perturbative corrections with a cutoff at low wavenumbers corresponding to the size of the box for a few points in Figure 3 and found that introducing this cutoff decreases both the linear variance $\sigma^2$ and the weakly nonlinear approximation of $\langle \delta^2 \rangle$, but their ratio also is decreased. Therefore the effect of such a cutoff is generally to decrease the perturbative values in Figure 3 so that the theoretical curve moves closer to the N-body results.

We can summarize this section as follows. All our N-body experiments support perturbation theory: nonlinear evolution makes $\langle \delta^2 \rangle$ larger than the linear variance, when $n = -2$. There is an opposite effect – a slow-down of growth rate, when $n > -1$, and the magnitude of the slow-down (the value of $I_2$, determined from N-body experiments) increases with $n$. The transition occurs at $n = -1$, when the nonlinear correction term is close to zero.

### 3.1.2 Comparison with the Hamilton ansatz

Hamilton et al. (1991) proposed a general formula relating the linear two-point correlation function of arbitrary shape and its strongly nonlinear counterpart. While physically motivated in its limits, the overall functional shape of the relation was modeled using N-body simulations of scale-free initial power spectra with $\Omega = 1$. A similar formula for power spectra was obtained by Peacock & Dodds (1994) and recently refined to take into account the dependence on the spectral index of the spectrum by Mo, Jain & White (1995). We will use this $n$-dependent formula to calculate the nonlinear variance and compare the result to the linear prediction as was done in the case of perturbative approximation and N-body simulations.

The ansatz for the relation between the linear power spectrum ($\Delta_L(k_0) = 4\pi k_0^3 P(k_0)$) and the nonlinear (evolved) one ($\Delta_E(k) = 4\pi k^3 P(k)$) is

$$\frac{\Delta_E(k)}{B(n)} = \Phi\left[\frac{\Delta_L(k_0)}{B(n)}\right] \tag{49}$$

where the wavevectors $k$ and $k_0$ are related by

$$k = [1 + \Delta_E(k)]^{1/3} k_0 \tag{50}$$

and $B(n)$ is a constant depending on the power spectrum index $n$. The values of $B(n)$ are 1.64, 1, 0.54 and 0.24 for $n = +1$, $n = 0$, $n = -1$ and $n = -2$ respectively. We use the fitting formula of Mo et al. (1995)

$$\Phi(x) = x \left(\frac{1 + 2x^2 - 0.6x^3 - 1.5x^{7/2} + x^4}{1 + 0.0037x^3}\right)^{1/2} \tag{51}$$

to calculate the nonlinear power spectrum $\Delta_E$ at specified values of $k_0$. Then, using equation (50) we find the value of $k$ corresponding to each of pairs $[k_0, \Delta_E(k_0)]$. The list of points $[k, \Delta_E(k)]$ is then fitted numerically by the linear power spectrum plus polynomial terms of higher order. This fitted shape of the nonlinear power spectrum $\Delta_E(k)$ is then used to calculate the nonlinear variance $<\delta^2>$. For simplicity we use a filter function in the form of a top-hat in Fourier space (that is we cut off the integration at some wavenumber $k_c$), the effect of which should be close to the effect of a Gaussian filter with the smoothing scale corresponding to $k_c$. The comparison of the results with the linear variance, $\sigma^2$, is shown in Figure 4. Given the numerous approximations applied in the calculations, the agreement between the results thus obtained with those of perturbative calculations and N-body simulations (Figures 2 and 3) is rather impressive.

## 3.2 The Peacock-Dodds spectrum

Scale-invariant spectra can be regarded as idealizations of "realistic" spectra, which arise either from a mixture of model assumptions and physics of the decoupling era, or from attempts to parametrize the observational data. Such spectra usually have slopes dependent on the wavenumber, $n(k) = d(\log P)/d(\log k)$, therefore the size and sign of weakly nonlinear corrections to $P(k)$ can be also expected to vary with $k$. As an example of such a spectrum we consider the following

$$\Delta(k) = \frac{(k/k_0)^\alpha}{1 + (k_c/k)^{\alpha-\beta}} \tag{52}$$

where $\Delta(k) = 4\pi k^3 P(k)/(2\pi)^3$. The parameters given by Peacock & Dodds (1994),

$$\begin{aligned}
k_0 &= 0.29 \pm 0.01 \ h \ \text{Mpc}^{-1} \\
k_c &= 0.039 \pm 0.002 \ h \ \text{Mpc}^{-1} \\
\alpha &= 1.50 \pm 0.03 \\
\beta &= 4.0 \pm 0.5,
\end{aligned} \tag{53}$$

were fitted to best match the data obtained from eight independent galaxy surveys. The reconstruction of this linear spectrum involved accounting for nonlinear evolution as well as redshift space distortions. If we accept the most probable value $\alpha - \beta = -2.5$ we may rewrite the spectrum in the following way

$$P(k) = \frac{Ck}{1 + (k/k_c)^n} \tag{54}$$

with $C = 2\pi^2 k_0^{-1.5} k_c^{-2.5}$ and $n = 2.5$. The spectrum of this kind was first considered by Peacock (1991) who quotes $k_c$ in the range $[0.015, 0.025] \ h \ \text{Mpc}^{-1}$ and $n = 2.4$.

Following equation (19), the linear variance of the density contrast averaged over a sphere of radius $R$ is

$$\sigma_R^2 = \frac{9\pi}{2k_0^{1.5}k_c^{2.5}R^4} \int_0^\infty \frac{\mathrm{d}k}{1+(\frac{k}{k_cR})^{2.5}} J_{3/2}^2(k) , \tag{55}$$

which is easily integrable numerically. Assuming the smoothing scale $R = 8\ h^{-1}$ Mpc we obtain

$$\sigma_8 = 0.69. \tag{56}$$

We calculated the nonlinear correction, using equations (27), (28) and (29) and a convenient approximation of a top-hat window function,

$$W_{TH}(kR) \approx \exp(-k^2R^2/9) , \tag{57}$$

which is an accurate representation of a real top-hat (eq. [17]) to within few percent in the entire range of integration. The convergence of the integrals is now ensured because for $k \to \infty$ the power spectrum (54) approaches $C/k^{1.5}$. The calculation must however be done for every considered smoothing scale independently as the spectrum is not scale-free. At $R = 8\ h^{-1}$ Mpc we find

$$I_2(8\ h^{-1}\text{Mpc}) = \frac{\langle \delta_2^2 + 2\,\delta_1\delta_3 \rangle_8}{\sigma_8^4} = 0.15 \tag{58}$$

and the value of $\sigma_8$ taking into account the weakly nonlinear corrections is

$$\sigma_{8,wnl} = 0.72 \tag{59}$$

which is less than 4 % above the linear value. This result is consistent with what we have shown above for the scale-free power spectra. At the scale of $8\ h^{-1}$ Mpc the spectrum has index close to $-1$ or slightly below this value. For such spectral indices we expect the weakly nonlinear variance to be equal to or slightly above the linear value.

One may ask how sensitive these results are to uncertainties in fitting parameters, given in equations (53). If we take the claimed error bars seriously, the resulting uncertainty in $I_2$ in equation (58) is rather small. Varying the least accurate parameter, $\beta$, from 3.5 to 4.5 gives $I_2 = 0.15 \pm 0.02$. Varying the other fitting parameters in the allowed ranges leads to negligible changes in $I_2$. Of course, we must keep in mind that the true uncertainties in the shape of $P(k)$ may be quite different. The parameters in equation (53) were not obtained from a homogeneous data set. Instead, the process of data reduction required a number of approximations, including a procedure applied to unify the data coming from different surveys and the method to account for the nonlinear evolution which is not expected to work equally well for different parts of the power spectrum.

## 3.3 Concluding remarks

We have computed the first nonlinear corrections to the variance of a density field undergoing gravitational instability. The results are confirmed by numerical simulations, and agree with the ansatz of Hamilton et al. (1991), as modified by Mo et al. (1995). There is indeed a previrialization effect, as conjectured by Peebles & Groth (1976) and Davis & Peebles (1977) and Peebles (1990). The effect depends on the initial power spectrum considered. For $n < -1$, the fluctuations grow faster than in linear theory, while for $n > -1$, we see "previrialization", or slow-down.

The magnitude of this effect grows when the relative amount of small-scale power is increased by increasing $n$. The effect is small at $n = -1$, corresponding to the transition between the slow-down and speed-up of the collapse. Our results can be used to explain the possible cause of differences of opinions about the size of the previrialization effect. Indeed, Evrard & Crone (1992) assumed $n = -1$ in their simulations, and saw no effect, while Peebles (1990) reached an opposite conclusion, using simulations with $n = 0$.

For scale-free power spectra the difference between the linear and weakly nonlinear approximation for the variance can be as high as 100% as in the case of the spectral index $n = -2$. For the realistic power spectrum of a class considered by Peacock & Dodds (1994) however, we have found that the correction induced by weakly nonlinear effects on $\sigma_8$ is very weak. This is purely by chance: the effective index of the realistic spectrum at the scale $R = 8h^{-1}$ Mpc happens to be close to $n = -1$. For such an index, as we have seen, the nonlinear correction is close to zero.

Our results are in agreement with previous studies of nonlinear interaction between perturbations at different scales. Using the fluid model for the evolution of structure Peebles (1987) found that for the initial power spectrum of index $n = 0$ (and some small scale cutoff) the smoothed standard deviation of the evolved field is smaller than its linear extrapolation. Weinberg & Cole (1992) performed a series of N-body simulations with Gaussian initial conditions and scale-free initial power spectra normalized so that the evolved $\sigma_8 = 1$. For the power spectrum index $n = -1$ they found that $\sigma_8$ grows at almost exactly the rate predicted by linear theory, while for $n = 0$ the required linear $\sigma_8$ was larger than the evolved value and for $n = -2$ smaller. Jain & Bertschinger (1994) found that for CDM spectrum normalized so that linear $\sigma_8 = 1$ the second-order effects increase $\sigma_8$ by 10%.

Why is the $n = -1$ index so special? At least a partial explanation may be provided by recalling the properties of the peculiar gravitational acceleration, $g$, known from linear perturbation theory. For a density field, smoothed on scale $R$, $g \propto R^{-(n+1)/2}$ (Peebles &

Groth 1976; Vittorio & Juszkiewicz 1987). If $n < -1$, $g$ diverges at large $R$: the peculiar acceleration field has an infinite coherence length (the same is true for the velocity field). The gravitational acceleration moves the fluid more or less uniformly, generating bulk flows rather than shearing motions. Such an acceleration field will move a collapsing mass concentration as a whole, without affecting its internal dynamics. Therefore, its collapse will be similar to that of an isolated spherical clump (in which $\delta$ grows faster than in linear theory). In the opposite regime, when $n > -1$, the dominant sources of acceleration are local, small-scale inhomogeneities. Hence, we can expect a slow-down of the growth rate: tidal effects will tend to generate nonradial motions and resist gravitational collapse. The above picture, based on linear theory arguments, can be made more rigorous by involving higher terms in the perturbative expansion and by studying the evolution of high density peaks rather than the evolution of regions, selected at random. This was recently done by Bernardeau (1994). The transition near $n = -1$ was also studied in N-body simulations by Klypin & Melott (1992) and Melott & Shandarin (1993), who concentrated on the properties of the peculiar velocity field.

## Acknowledgements

We wish to thank our referee, Jim Peebles, for important comments and suggestions. We acknowledge enjoyable and useful discussions with Francis Bernardeau, Michał Chodorowski, Paolo Catelan, Josh Frieman and Román Scoccimarro. We are grateful to David Weinberg for kindly providing us with his N-body simulations. The computational means (CRAY-98) have been made available to us thanks to the scientific council of the Institut du Développement et des Resources en Informatique Scientifique (IDRIS). We thank Alain Omont and James S. Langer for their hospitality at Institut d'Astrophysique de Paris and at ITP at Santa Barbara, respectively. This research has been supported in part by the French Ministry of Research and Technology within the programme RFR, Polish KBN grants No. 2P30401607 and 2P30401707, and a NSF grant No. PHY94-07194.

# Figure captions

**Figure 1** Weakly nonlinear corrections to the variance $\langle \delta_2^2 \rangle + 2 \langle \delta_1 \delta_3 \rangle$, divided by $\sigma_G^4$ (eq. [30]) for a Gaussian filter and different scale-free power spectra. Symbols show the results of numerical integration up to different values of $k_c R$, where $k_c$ is the cutoff wavenumber in the initial spectrum and $R$ is the scale of the final smoothing. Only for $n = -2$ the integration converges to 0.86, which is slightly different from the value obtained analytically (eq. [37]). In the limit of $k_c R \to 0$ all curves converge to values corresponding to the case of no smoothing (eq. [33]).

**Figure 2** The ratio of the nonlinear to linear variance, $\langle \delta^2 \rangle / \sigma^2$, as a function of $\sigma^2$ for different initial power spectra. Open symbols correspond to the results of N-body simulations with Gaussian smoothing lengths of $L/12.5$ (squares), $L/25$ (triangles) and $L/50$ (circles), where $L$ is the size of the simulation box. Filled triangles show the results of perturbative calculations with final smoothing radius $L/25$ and the small scale cutoff $r_{Ny}/4$ for $n = -1$, $r_{Ny}/2$ for $n = 0$ and $r_{Ny}$ for $n = +1$.

**Figure 3** The comparison of the perturbative vs. N-body results for the power spectrum of index $n = -2$ and top-hat smoothing. Open symbols show the ratio of the nonlinear to linear variance, measured in N-body experiment at different expansion factors and eight final smoothing scales starting from $R = 10^{-2.2}$ times the size of the simulation box and spaced by 0.2 in log. The corresponding perturbative results (filled symbols) were calculated assuming the normalization of the simulated initial spectrum and using equation (38).

**Figure 4** The ratio of the nonlinear to linear variance, as a function of $\sigma^2$ for different initial scale-free power spectra. The nonlinear variance was obtained from the nonlinear power spectrum given by an ansatz, similar to the one, proposed by Hamilton et al. (eqs. [49] – [51]). Both linear and nonlinear values were calculated using a top-hat filter in Fourier space.

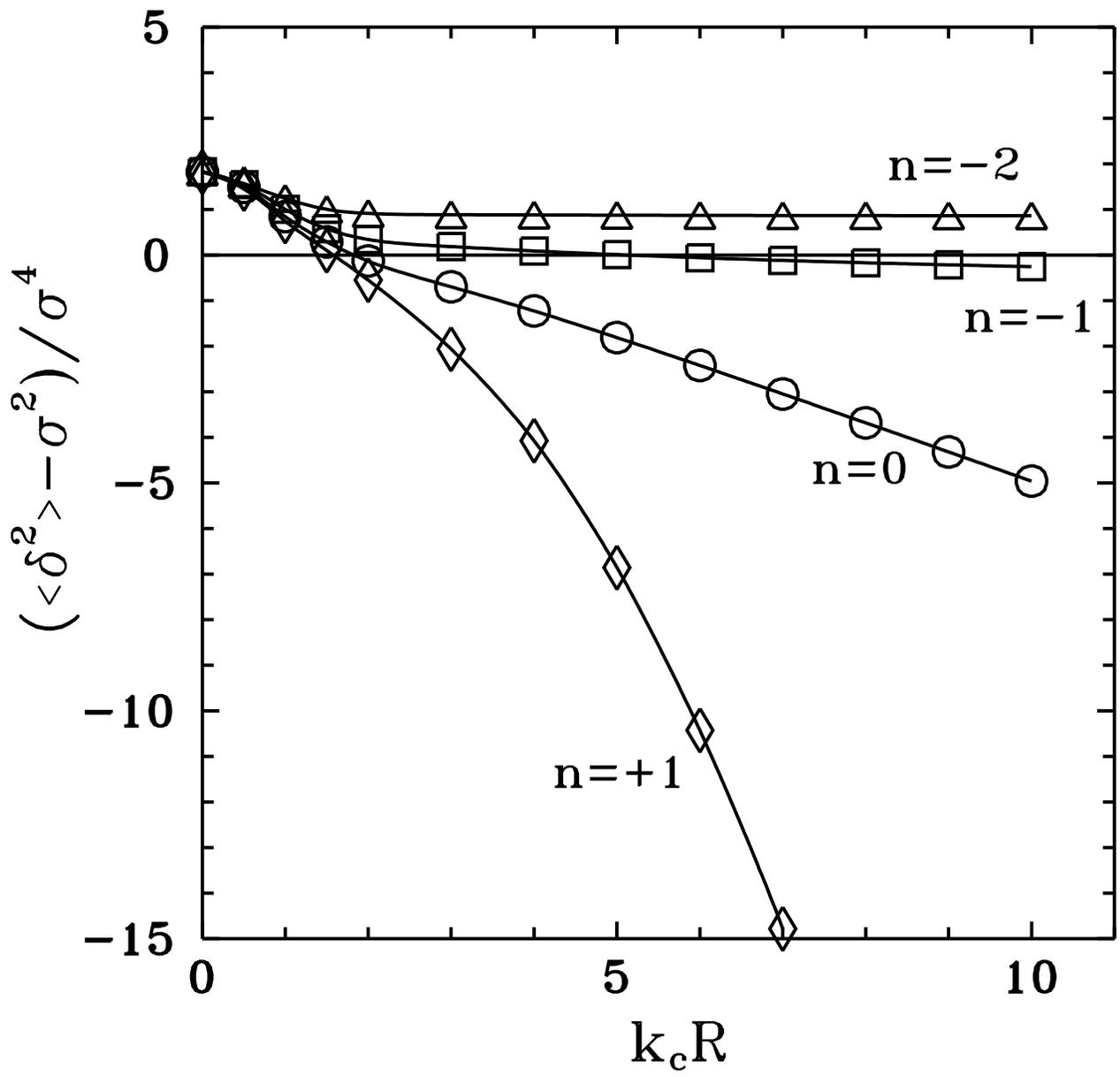

Figure 1

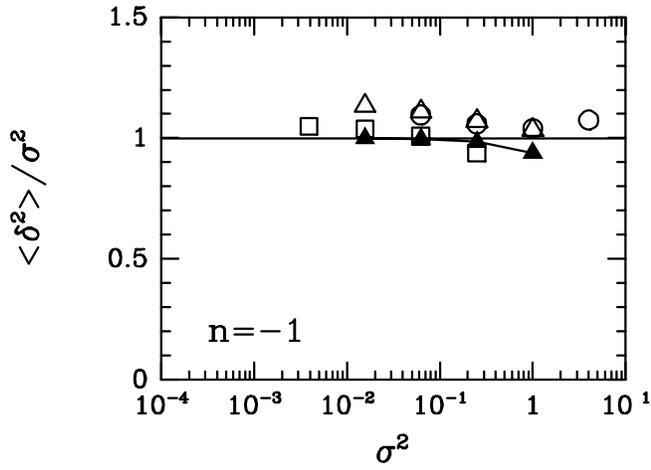

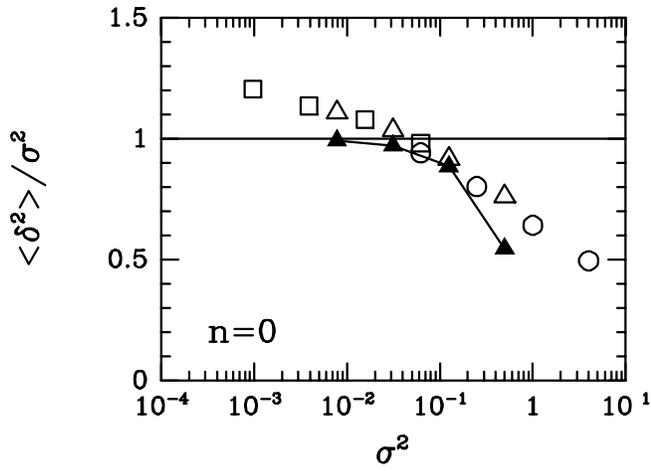

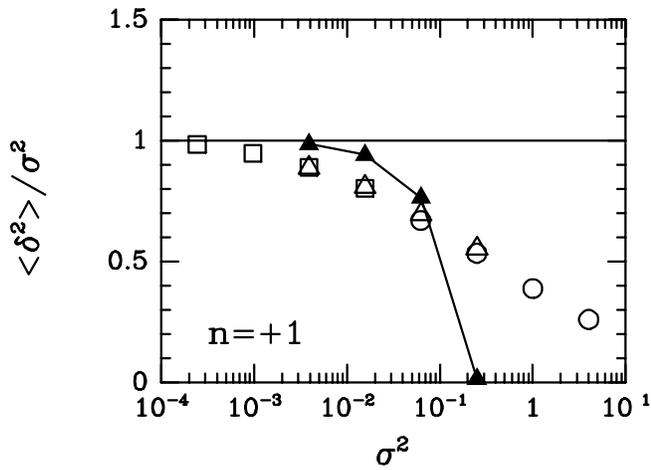

Figure 2

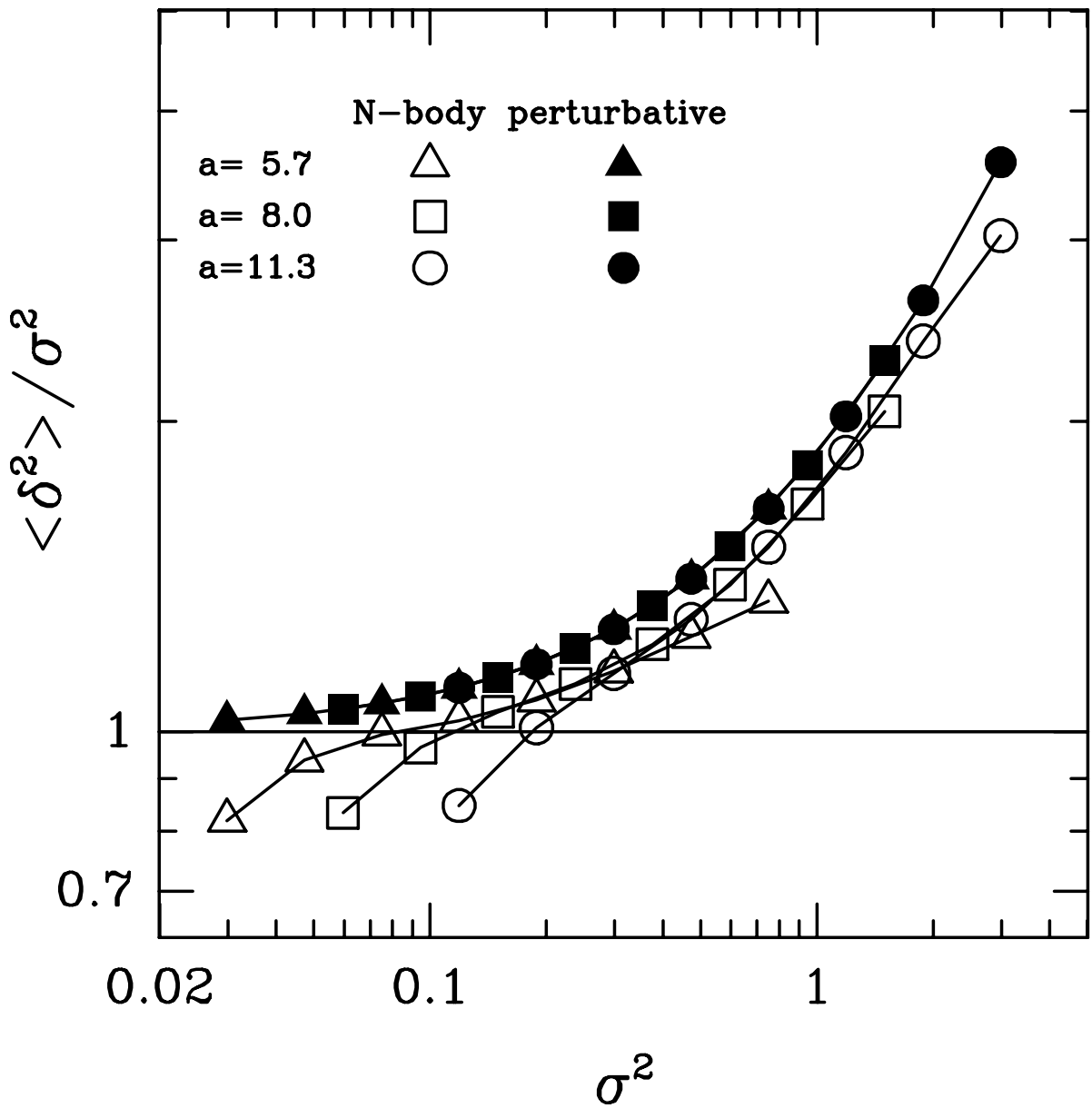

Figure 3

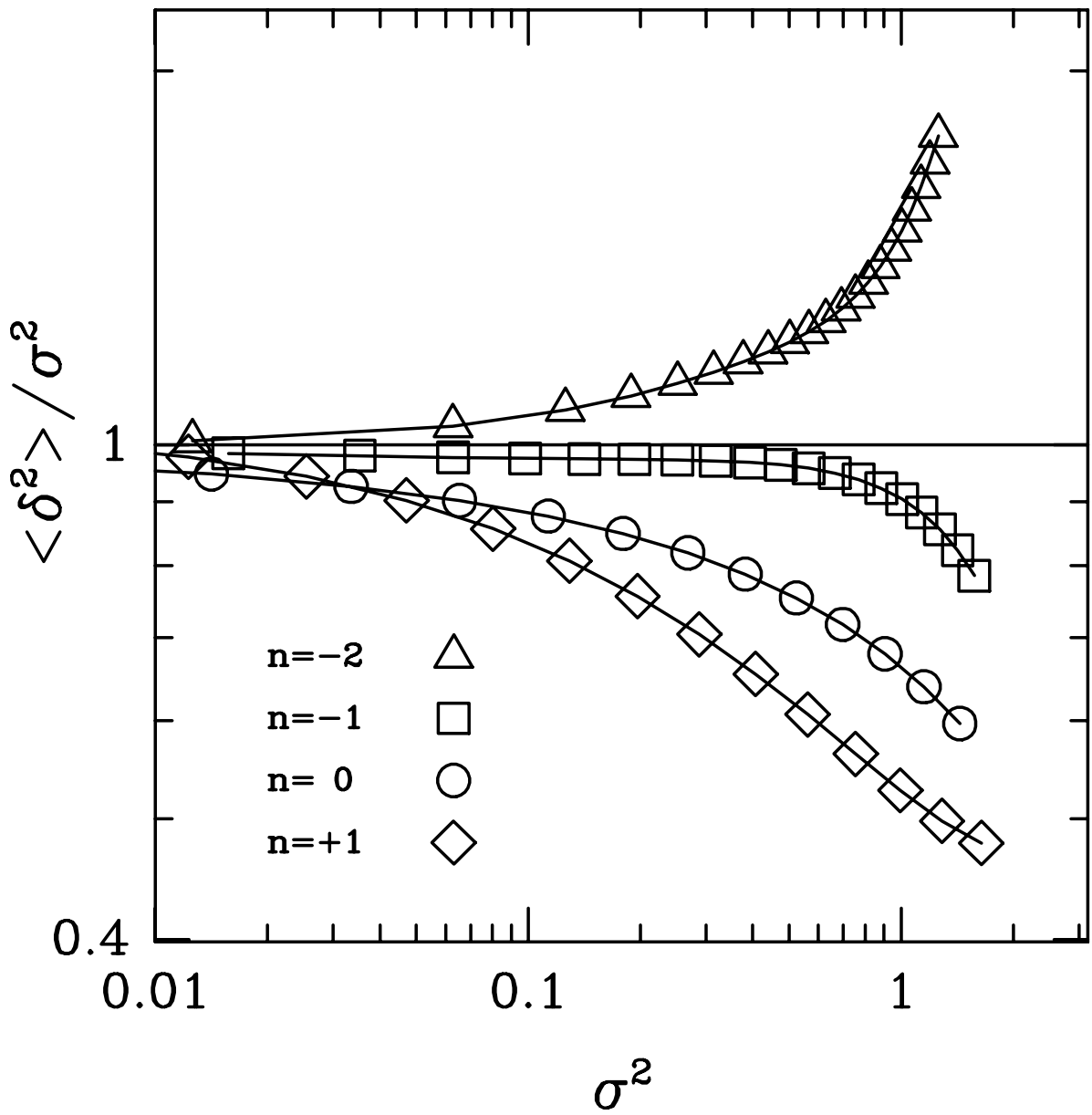

Figure 4